\documentclass[a4paper,11pt]{article}

\usepackage{jinstpub} 
\usepackage{comment}
\usepackage{subcaption}

\title{\boldmath Radon emanation rate measurements using liquid scintillation counting}

\author[a,1]{A. B. M. R. Sazzad,\note{Corresponding author.}}
\author[a]{P. Acharya,}
\author[a]{P. Back,}
\author[a]{J. Busenitz,}
\author[a]{D. Chernyak,}
\author[a,2]{Y. Meng,\note{Now at Shanghai Jiao Tong University, China}}
\author[a]{A. Piepke,}
\author[a,3]{C. A. Rhyne\note{Now at Adelphi Technology, Inc.}}
\author[a,4]{R. Tsang\note{Now at Canon Medical Research USA, Inc.}}

\affiliation[a]{Department of Physics and Astronomy, University of Alabama, 514 University Blvd., Tuscaloosa AL 35487, USA}

\emailAdd{asazzad@crimson.ua.edu}

\abstract{This article describes a radon emanation measurement technique using liquid scintillator counting. A model for radon loading and transport is described, along with its calibration. Detector background and blank have been studied and quantified. The Minimal detectable activity has been determined for the counting setup using a toy Monte Carlo simulation. The measurement technique is validated using a butyl rubber sample previously used for cross-calibration between different radon counting facilities.}


\begin{document}
\maketitle
\flushbottom

\section{Introduction}
\label{sec:intro}

Radon and its progeny are important sources of background events in low-energy rare-event-searches, such as experiments studying solar neutrinos, double beta decay, and dark matter. 
Being a noble gas with a mean lifetime of $\rm \tau_{Rn}=5.5143\; d$, $^{222}$Rn can travel relatively large distances before decaying into chemically more active, immobile nuclides. 
Even within large detectors, radon can diffuse from external components toward the sensitive region, particularly through gas-filled voids in the shielding. 
It can diffuse through membranes~\cite{Schnee_Varies_Bowles_2023, WOJCIK2000158, MENG2020108963} and constantly emanates in trace amounts from uranium-containing detector components. 
Radon released from materials is driven by nuclear recoil and diffusion. 
Radon emanation from these components contributes to the overall background rate in low background experiments and must be accounted for in modeling. 

Different techniques are being used for the measurement of radon at low concentrations, typically by detecting its $\alpha$-decay or the various other decay modes of its progeny~\cite{WANG_1999,Zuzel_Simgen_2009,Rumleskie_2015,Nakano_2017,Aprile_XENON1T_2021,Pelczar_2021}. 
Here, we describe a liquid scintillator-based detector setup using time-correlated decays of the radon progeny $^{214}$Bi and $^{214}$Po. 
This approach utilizes the good solubility of radon in aromatic and long-chain aliphatic organic materials, as found in liquid scintillator. 
The approach is simple and cost-effective. 
The setup, and methods similar to what is described here, have been used during the preparation of the LZ dark matter search, with results reported in reference~\cite{LZradioactivityPaper}. 
This paper gives a detailed account of the technical and data analysis approach and improvements made to it.

\section{Radon emanation setup and procedure}
\label{sec:setup}

The radon emanation measurement technique we are describing consists of three principal steps: 
\begin{enumerate}
    \item accumulating the $^{222}$Rn emanating from a sample in a sealed collector chamber, 
    \item loading of the collected $^{222}$Rn into liquid scintillator (LS), and 
    \item counting the delayed coincidences resulting from the \textsuperscript{214}Bi--\textsuperscript{214}Po $\mathrm{\beta-\alpha}$ decay sequence using the liquid scintillator. 
\end{enumerate}
The subsequent sections describe all steps of this procedure in detail. 

\subsection{Liquid scintillator}
The liquid scintillator (LS) used for this work is prepared in the lab and then aged in air-tight containers to allow the decay of radon dissolved in it during preparation and handling.
The aging lasts for at least 4 weeks after mixing, ensuring that less than 0.6\% of the initially dissolved radon remains. 
The LS consists of 80v\% dodecane, 20v\% pseudocumene (1,2,4-trimethylbenzene) and 1.5 g/L PPO (2,5-diphenyloxazole). The primary scintillator, pseudocumene, is diluted with dodecane to prevent the cocktail from damaging or grazing the acrylic counting cells.

\subsection{Radon accumulation and loading}
\begin{figure}
    \centering
    \includegraphics[width=\textwidth]{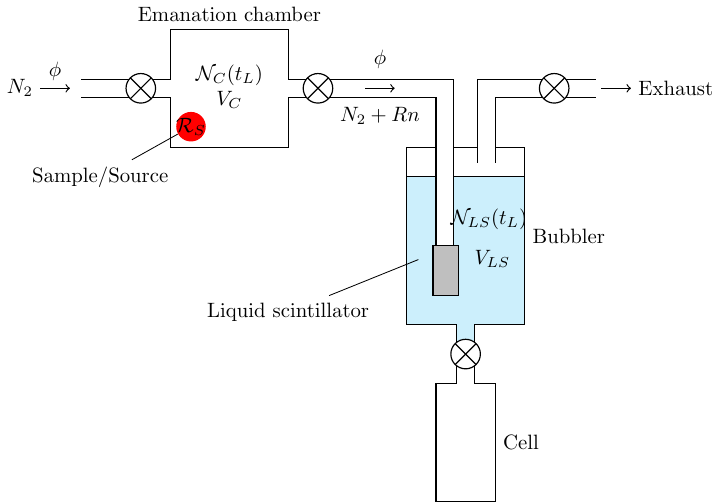}
    \caption{Schematic of the emanation setup (not to scale), showing one collector chamber. A nitrogen gas stream is used to transport radon, accumulated in the emanation chamber, into the LS, contained in the bubbler. After the gas flow ends, the LS is drained into the acrylic counting cell at its bottom. 
    }
    \label{fig:emanationSystem}
\end{figure}
The radon emanation measurement process consists of three key steps. First, samples are cleaned with acetone, methanol, or isopropyl alcohol to ensure that measurements reflect the bulk outgassing properties rather than surface contamination. The samples are then placed in one of two independently functioning, 2.6 L electropolished stainless steel emanation chambers. To minimize external radon contamination, the chambers are evacuated and back-filled with boil-off nitrogen gas. Radon gas emitted from the samples is accumulated over a period of two weeks or more, achieving 92\% of the maximally possible radon activity.

Once the accumulation phase is complete, the boil-off nitrogen is flowed through the emanation chamber to transport the radon into LS contained in a metal bottle. A metal sparger, at the bottom of the bottle, breaks the gas into small bubbles, maximizing surface contact and enhancing radon transfer to the LS. The radon transfer efficiency from the gas phase into the liquid has been estimated to exceed 80\%, based on tests using a series of Nalgene lab bottles and subsequent radon content measurement via germanium detector. This value does not enter into the data analysis. 

After the transfer, the LS—now containing the radon—is drained into cylindrical acrylic cells ($\varnothing$=110 mm, height 32.6 mm ) to detect decays from radon progeny, specifically $^{214}$Bi and $^{214}$Po. 
On average, 150 ml of LS is transferred into a single counting cell. Around 87\% of the LS containing the radon is successfully moved. The cells are designed with filling valves to limit air exposure and prevent environmental radon contamination. Each measurement uses a new cell to avoid residual radon carryover. Additionally, a mass flow meter ensures that a consistent LS volume is used for each assay, preventing overfilling or spills into the gas handling system.

\subsection{Coincidence counting}\label{set_up}
The counting cell is equipped with a photomultiplier tube (PMT) and installed in the passive lead shield, of one of two ``counting setups''.
These setups utilize distinct data acquisition (DAQ) systems and are equipped with different PMTs, PMT holders, and passive shields. 

Counting of the time-correlated decays of the $^{222}$Rn progeny $^{214}$Bi and $^{214}$Po (called Bi-Po events), growing into secular equilibrium with its parent after loading, is used to quantify the radon content of the LS and with it the outgassing rate of the sample. 
Because of the short mean lifetime of $^{214}$Po ($\rm \tau_{Po}=236.0\; \mu s$), delayed coincidences offer a powerful way to reduce random backgrounds unrelated to radon.

To read out the scintillation signal, the acrylic counting cell is optically coupled at one of its flat sides to a 3" Hamamatsu R1307 low radioactivity PMT using Saint-Gobain Crystals BC-630 optical grease to ensure optimal light transmission.
Care is taken to exclude air bubbles that would lead to light loss and degraded resolution. 
To enhance photon collection, the cell mantle is covered with a Teflon reflector, which has been shown to nearly double the light collection. This improvement proved useful when interpreting $\gamma$-source calibration data.
The entire PMT-cell assembly is encased in a lead-shield for low-background counting.  

Two independent DAQ systems are used to read, digitize, and store the counting data. One of them is based on multiple Nuclear Instrumentation Modules (NIM), and the other utilizes a waveform digitizer.
For both setups, the data are saved in the form of ROOT trees. Customized ROOT scripts have been developed to perform event selection and data analysis.

The first setup (``coincidence setup'') uses a DAQ system based on NIM and Computer Automated Measurement and Control (CAMAC) modules. It utilizes discriminators, logic modules, analog-to-digital converters (ADCs), and a scaler-pulse generator combination. 
Pulse pairs detected within 2500 $\mu$s of each other, corresponding to 10.6 $^{214}$Po mean lifetimes, are digitized and saved for data analysis. This long correlation time allows the measurement of the random coincidence background together with the signal.
To minimize data size, only event pairs are saved, with the initiating event labeled as the $\beta$-like event and the coincident trailing event labeled as the $\alpha$-like event in separate branches of the ROOT TTree structure.

The second counting setup (``digitizer setup'') makes use of a  CAEN DT5730 waveform digitizer. The digitizer is operated in constant fraction discriminator mode. We are currently using the CAEN-provided CoMPASS software to save PMT pulse areas and timestamps in a ROOT tree. A ROOT script is used to select sequential event pairs occurring within a 2500 $\mu$s time window for further analysis. 
We are considering implementing off-line pulse shape fitting, which would greatly increase the amount of data collected by the waveform digitizer.

The system dead time was investigated employing a Wavetek model 145 function generator. PMT-like pulses were fed into the coincidence DAQ with different frequencies. The percentage loss of the detected pulse was plotted against the frequency. Because generator pulses are evenly distributed in time, the loss relation has the form of a step function, jumping from 0 to 50\% when every other pulse is lost due to dead time. This step was observed at 4309 Hz, indicating a processing time of 232 $\rm \mu$s per event. 

Our measurements typically involve rather moderate counting rates. The highest rate point in the lower panel of figure~\ref{fig:optimalCut}, a calibration data set, has an estimated dead time of 0.17\%, which is small enough to be neglected in the analysis.

\section{Radon loading model}
\label{sec:model}

\subsection{Model development} 
Our assay method requires the accumulation of radon in a collection chamber, transfer by gas stream into a bubbler, solution in LS, and filling into a counting cell. The amount of harvested radon and various losses need to be understood and modelled. This section describes our radon transport model and how it has been tuned with data obtained using a calibrated Pylon model RN-1025 \textsuperscript{222}Rn source. 

Let $\mathcal{R}_{\rm S}$\footnote{We use the cursive letter $\mathcal{R}$ to denote radon outgassing rates. Block script $\rm R$ describes event rates.} be the rate at which a sample releases radon into the surrounding gas. $\mathcal{R}_{\rm S}$ is the quantity of interest. During the growth period, with a sample present, the number of radon atoms contained in the emanation chamber at emanation time $\rm t_E$, $\rm \mathcal{ N}_C(t_E)$\footnote{$\rm \mathcal{N}$ denotes a number of atoms, block script $\rm N$, depending on context, a number of decays or events.}, is given by:
\begin{equation}
    \mathrm{
    \frac{\partial \mathcal{N}_C(t_E)}{\partial t_E} = \mathcal{R_S} - \frac{\mathcal{N}_C(t_E)}{\tau_{Rn}}.} \label{eqn:growth}
\end{equation}

Assuming no radon to be present at $\rm t_E=0$ as a boundary condition,
\begin{equation}
\mathrm {
\mathcal{N}_C(t_E)=\mathcal{R_S}\cdot \tau_{Rn}\cdot \left( 1-e^{-t_E/\tau_{Rn}} \right)}.\label{eq:accumulation}
\end{equation}

In a first transfer step, the radon, accumulated in the emanation chamber is flushed into the bubbler by means of a metered nitrogen gas stream. The radon removal from the nitrogen-filled collection chamber, thus, involves dilution. We are modelling this step assuming the removed radon fraction to be proportional to the ratio of the volume of the emanation chamber and that of the gas used in the transfer. Denoting the gas flow rate as $\rm \phi$ and the emanation chamber volume with $\rm V_C$, the instantaneous transfer rate is taken to be proportional to $\rm \frac{\phi}{V_C}$. 

The $\rm \frac{\phi}{V_C}$-dependence of the radon transfer has been verified in a stand-alone experiment. Using the Pylon source, a $\rm 5800\; cm^3$ sealed container was filled with nitrogen, containing radon. The box was then purged with clean nitrogen, via a metered gas flow controller, the same way the emanation chamber is unloaded. The exiting gas stream was routed through a RAD7 detector, and the time dependence of the exiting radon activity was determined. This test resulted in a ratio of modeled over measured time constant of $0.955\pm 0.052$, confirming this aspect of the radon transport model.

Starting the gas flow constitutes a discontinuous change, starting the loading time counting $\rm t_L$.
The $\rm \frac{\phi}{V_C}$ related loss term describes the radon removal from the emanation chamber: 
\begin{equation}
\mathrm{
   \frac{\partial \mathcal{N}_{C}(t_L)}{\partial t_L}  =  
    \mathcal{R}_S - \frac{\phi}{V_{C}} \cdot \mathcal{N}_C(t_L) - \frac{\mathcal{N}_C(t_L)}{\tau_{Rn}}
    }
    \label{eqn:srcRate}
\end{equation}

For sample measurements $\mathcal{R}_{\rm S}$ is small. However, this is not true for measurements with the Pylon radon source present. 
The start of radon transfer is taken as time zero ($\rm t_L=0$). 
For runs with a sample, the amount of radon present in the chamber at time zero is given by equation~\ref{eq:accumulation}, serving as a boundary condition connecting the two sequential phases.
For runs with the Pylon flow-through source, $\mathcal{R}_{\rm S}$ and its uncertainty are known and taken from the calibration certificate of the source.

The radon removed from the emanation chamber is transferred to the LS with volume $\rm V_{LS}$.
The rate of change of the number of radon atoms in the liquid scintillator, $\rm \mathcal{N}_{LS}(t_L)$, has a gain, equalling the loss of the emanation chamber, and a loss term due to desorption of radon because of the gas flow, plus a small decay term. 
The existence of the desorption term has been verified by loading LS with radon, determining its activity by means of low background germanium spectrometry, bubbling with clean nitrogen gas, followed by counting.
The balance of absorption and desorption of radon atoms in the LS is described by:
\begin{equation}
    \mathrm{
    \frac{\partial \mathcal{N}_{LS}(t_L)}{\partial t_L}
    }  =  
    \mathrm{
    \alpha \cdot \frac{\phi}{V_C} \cdot \mathcal{N}_C(t_L) - \beta \cdot \frac{\phi}{V_{LS}} \cdot \mathcal{N}_{LS}(t_L) - \frac{\mathcal{N}_{LS}(t_L)}{\tau_{Rn}},
    } 
    \label{eqn:bubblerRate}
\end{equation}
where $\mathrm{\alpha}$ and $\mathrm{\beta}$ are interpreted as dimensionless absorption and desorption coefficients, respectively.

The transfer of radon through both devices is described by coupling these equations. The solution of equation~\ref{eqn:srcRate} is substituted into equation~\ref{eqn:bubblerRate} to obtain the transfer function. The number of radon atoms in the liquid scintillator, after flowing gas for a period $\rm t_L$, is: 
\begin{eqnarray}
    \mathrm{
    \mathcal{N}_{LS}(t_L) 
    }
    & = & 
    \mathrm{
    A_{LS}(t_L)\cdot \tau_{Rn} = \alpha \cdot \mathcal{R}_S\cdot \frac{\phi}{V_C} \cdot \left[ \tau_1\cdot \tau_2 
    - \frac{\tau_1 \cdot e^{-t_L/\tau_1}}{\frac{1}{\tau_2}-\frac{1}{\tau_1}}
    + \frac{\tau_2 \cdot e^{-t_L/\tau_2}}{\frac{1}{\tau_2}-\frac{1}{\tau_1}}
    \right.
    } \nonumber \\
      &  &
      \mathrm{
      \left. + 
     \frac{\tau_{Rn} \cdot (1-e^{-t_E/\tau_{Rn}})}{\frac{1}{\tau_2}-\frac{1}{\tau_1}} \cdot
     \left( e^{-t_L/\tau_1} -e^{-t_L/\tau_2}  \right)
    \right]
    \label{eqn:Nls}
    }
\end{eqnarray}
To obtain a more compact solution we defined two variables: $\rm \frac{1}{\tau_1}=\frac{\phi}{V_C} + \frac{1}{\tau_{Rn}}$ and $\rm \frac{1}{\tau_2}=\beta \cdot \frac{\phi}{V_{LS}} + \frac{1}{\tau_{Rn}}$.
The model contains two unknown parameters, $\alpha$ and $\beta$, determined experimentally using the calibrated Pylon \textsuperscript{222}Rn source. For source runs $\rm t_E=0$.
Note that equation~\ref{eqn:Nls} depends implicitly on the model parameter $\beta$ through its variable $\tau_2$.

\subsection{Determination of absorption and desorption coefficients} 
A suite of radon-spiked liquid scintillator samples with varying loading times was counted to determine the two unknown parameters $\alpha$ and $\beta$ of the loading model. 

Known amounts of \textsuperscript{222}Rn, obtained from the calibrated Pylon source, were loaded into liquid scintillator using the procedure described before. 
Care was taken to use the same $\rm V_{LS}$ for all measurements.  After transfer into a counting cell, the activities of the $\gamma$-ray emitting radon progeny $^{214}$Pb and $^{214}$Bi were determined using a low background germanium detector. This data allowed to infer the radon activity, after correction for growth and decay. The activity determination by means of a Ge detector relies on its GEANT4-based acceptance model that has previously been verified using calibrated sources~\cite{TSANG201975}. This approach decouples the calibration of the radon loading model from the estimation of the delayed coincidence cut efficiencies, described in section~\ref{sec:optimizing}.

Only data collected 24 hrs after the end of loading were included in the Ge data analysis to ensure the secular equilibrium of the progeny had been reached. The determination of the radon content of the liquid scintillator at loading time uses the variance-weighted average of six peak-wise activity values. A peak-wise 3\% systematic uncertainty has been added. 

\begin{figure}
     \centering
     \begin{subfigure}{0.49\textwidth}
         \centering
         \includegraphics[width=\textwidth]{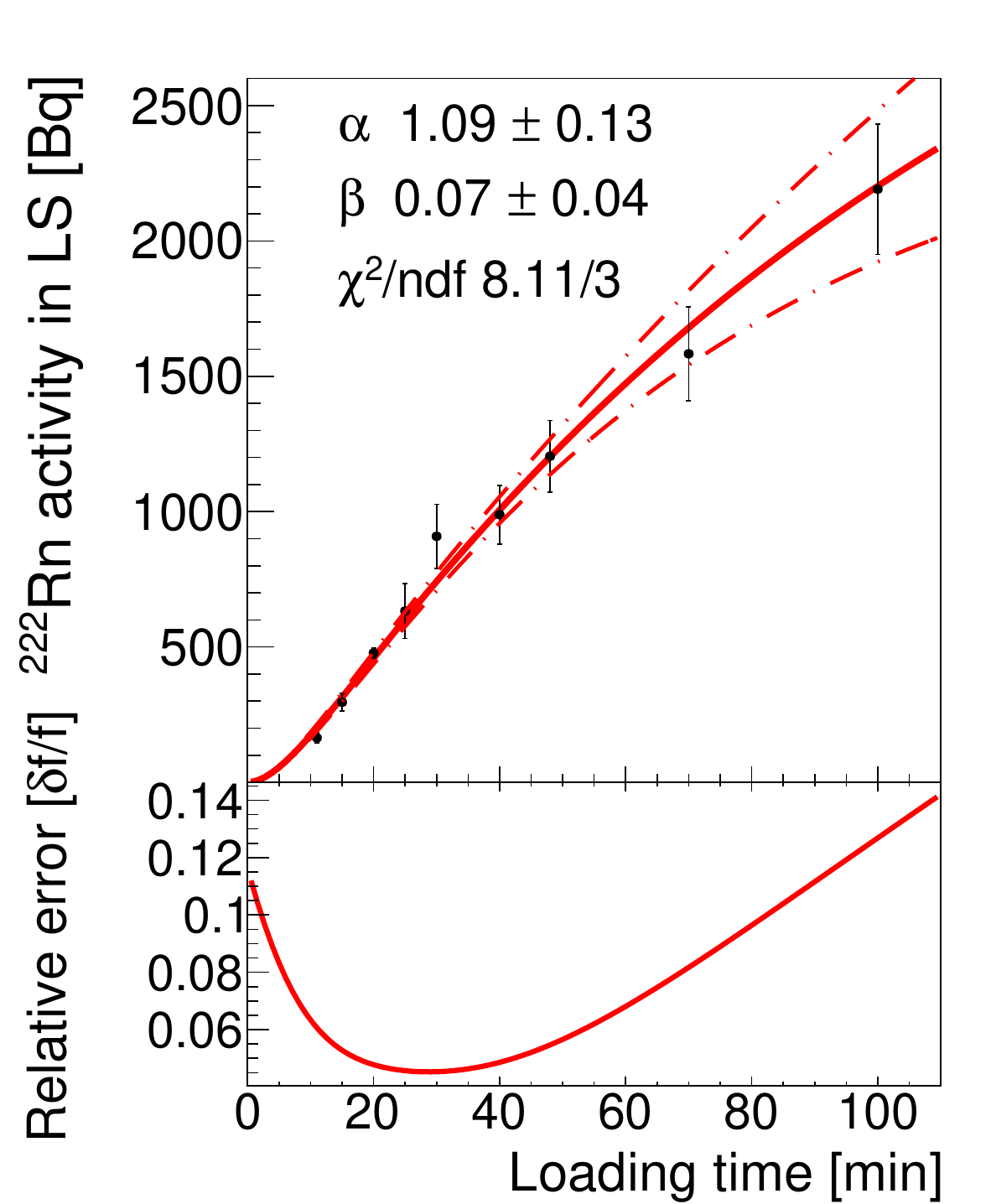}
         \label{fig:srcLoading}
     \end{subfigure}
     \begin{subfigure}{0.49\textwidth}
         \centering
         \includegraphics[width=\textwidth]{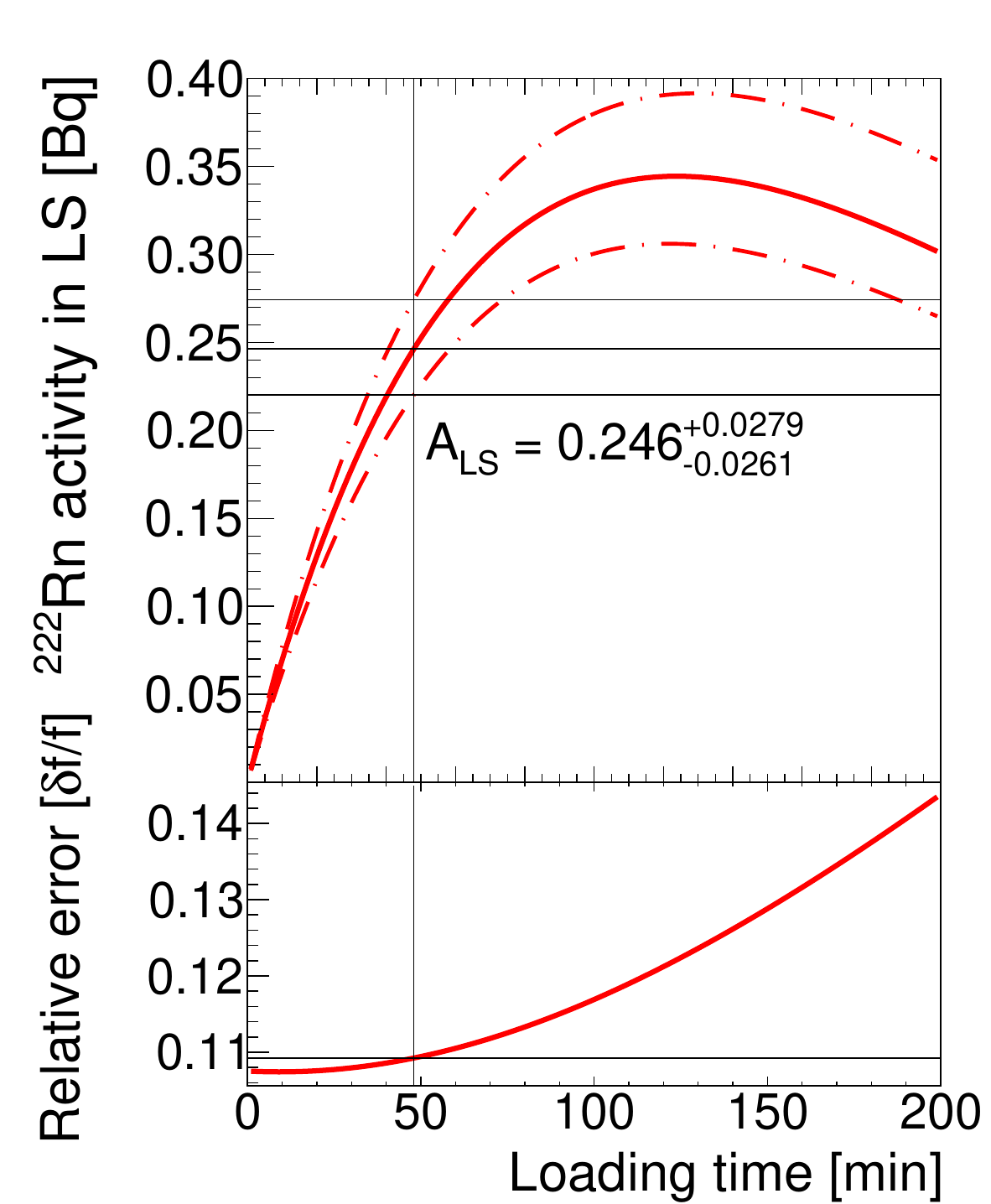}
         \label{fig:relativeErrorSrc}
     \end{subfigure}
     
     \caption{Left: $\rm t_L$-dependent \textsuperscript{222}Rn 
    activity in LS after loading with the Pylon source.   
    The solid line shows the fit of the data to equation~\ref{eqn:Nls} with $\alpha$ and $\beta$ free-floating, the dash-dotted line the uncertainty band. The relative uncertainty for this operation mode is also depicted in the bottom panel. \\
     Right: calculated LS \textsuperscript{222}Rn activity for $\mathrm{\mathcal{R}_S} = 1\; \mathrm{s^{-1}}$, using $\alpha$ and $\beta$ derived from the calibration data. This curve is based on $\mathrm{V_C}$ instead of $\rm V_S$. The relative uncertainty for this operation mode, as shown in the bottom panel, indicates that a 48-minute loading time maintains the relative uncertainty just below 11\%. \\
     }
     \label{fig:loadingModel}     
\end{figure}

Figure \ref{fig:loadingModel} (left) depicts the radon activity of the source-loaded LS, observed for different loading times. These data are fitted with the loading model, given by equation \ref{eqn:Nls}, converting from radon atoms to $\rm A_{LS}(t_L)$. $\mathrm{\chi^2}$ minimization is used (implemented via TMinuit in ROOT) to determine $\alpha$ and $\beta$. 
The parameters $\mathrm{\mathcal{R}_S, V_S, V_{LS}}$, $\mathrm{\phi}$ and their uncertainties are used as Gaussian penalty terms during the minimization. For Pylon radon source runs, its internal volume $\rm V_S$, taken from the manual, replaces $\mathrm{V_C}$ in equation \ref{eqn:Nls}.
The error estimate of the loading efficiency, indicated by the dash-dotted bands in figure~\ref{fig:loadingModel}, therefore, includes the uncertainties of these parameters. The uncertainty was estimated from the fit-derived covariance matrix to account for parameter correlations.

The right side of figure~\ref{fig:loadingModel} shows the loading model, applied to the emanation chamber, now using $\rm V_C$ instead of $\rm V_S$.
The differences in time dependence (for equal $\alpha$ and $\beta$ values), compared to the calibration data shown on the left, stem from the quite different internal volumes. The source manual gives its internal volume as $\rm V_S=106\pm 12\; cm^3$. The estimated volume of the tubes, connecting the source and the bubbler, is used as the volume uncertainty. The gas-filled volume of the emanation chamber, on the other hand, is $\rm V_C=2650\pm 5\; cm^3$, 25 times larger than what is used for the calibration of the loading model. 
We rely on the loading model to properly describe this difference.

$\mathrm{V_{LS}}$, is estimated from the amount of scintillator transferred into the acrylic cell plus what remains in the bubbler. 
Averaged over all radon-spiked measurements, we obtain $\rm V_{LS}=174.6\pm 4.4\; ml$ (RMS stated).
$\mathrm{\phi}$ is obtained from a gas flow controller, with a manufacturer-stated uncertainty of 1\% of its full-scale reading.
The emanation and loading times ($\rm t_E$ and $\rm t_L$) are considered to have negligible uncertainty and carry no associated penalty terms.

The relative uncertainties of the model-derived \textsuperscript{222}Rn activities of the LS are shown in the bottom panels of figure~\ref{fig:loadingModel}, increasing with loading time. 
To minimize this uncertainty and to ensure a substantial transfer of \textsuperscript{222}Rn from the carrier gas to the scintillator, a loading duration of 48 minutes has been adopted.
Figure \ref{fig:loadingModel} (right) shows the bubbling time dependence of the \textsuperscript{222}Rn activity recovered from the emanation chamber into the liquid scintillator, assuming a sample emanation rate of $\mathrm{\mathcal{R}} = 1\; \mathrm{s^{-1}}$. 
Because of the normalization to one, the curve can be interpreted as the radon recovery and transfer fraction.

\section{Bi-Po coincidence event acceptance}
\label{sec:optimizing}

After loading into the LS, the harvested $\rm ^{222}$Rn decays. We denote the decay time counting with $\rm t_C$, with zero chosen to be the end of loading.
To determine $\mathcal{R}_S$, we observe the $\rm t_C$-dependence of the LS activity, $\rm A_{LS}(t_C)$, using $^{214}$Bi-$^{214}$Po delayed coincidences. 
To ensure secular equilibrium has been established between $^{222}$Rn and $^{214}$Bi ($\rm \tau_{Bi}=28.44\; min$), data taking starts no sooner than 3 hr after the end of loading. In secular equilibrium, the decline of $\rm A_{LS}(t_C)$ follows a simple exponential, determined by $\tau_{Rn}$. The counting data is summarized by a single number, $\rm A_{LS,0}$, the fitted activity at $\rm t_C=0$. 

For both counting setups, introduced in section~\ref{set_up},
event selection is based on cuts on the event attributes ``$\beta$-like'' (prompt) energy deposit, and ``$\alpha$-like'' (delayed) energy deposit.

\subsection{Energy calibration of the setup}
Cuts placed on the prompt and delayed energy deposits provide selectivity in the event selection. Shifts in detector gain, therefore, impact the cut efficiency and with it the radon detection efficiency.
To control gain variations, energy calibration measurements with $\gamma$-ray emitting point sources are performed before and after each Bi-Po coincidence run. 
This is done by means of Compton edges visible in the energy spectra.

\begin{figure}
    \centering
    \begin{subfigure}{0.49\textwidth}
        \centering
        \includegraphics[width=\textwidth]{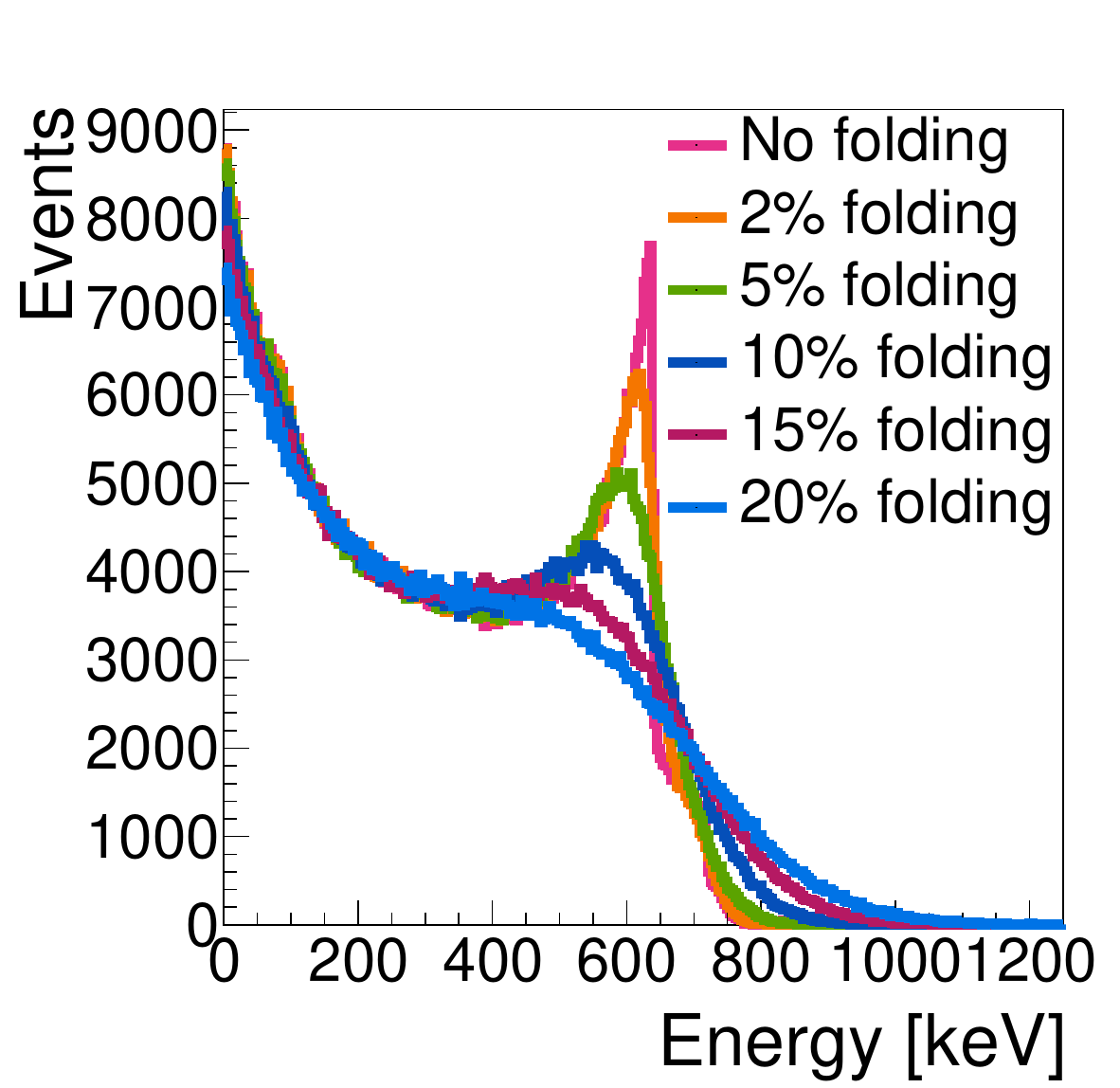}\\
    \end{subfigure}
    \begin{subfigure}{0.49\textwidth}
        \centering
        \includegraphics[width=\textwidth]{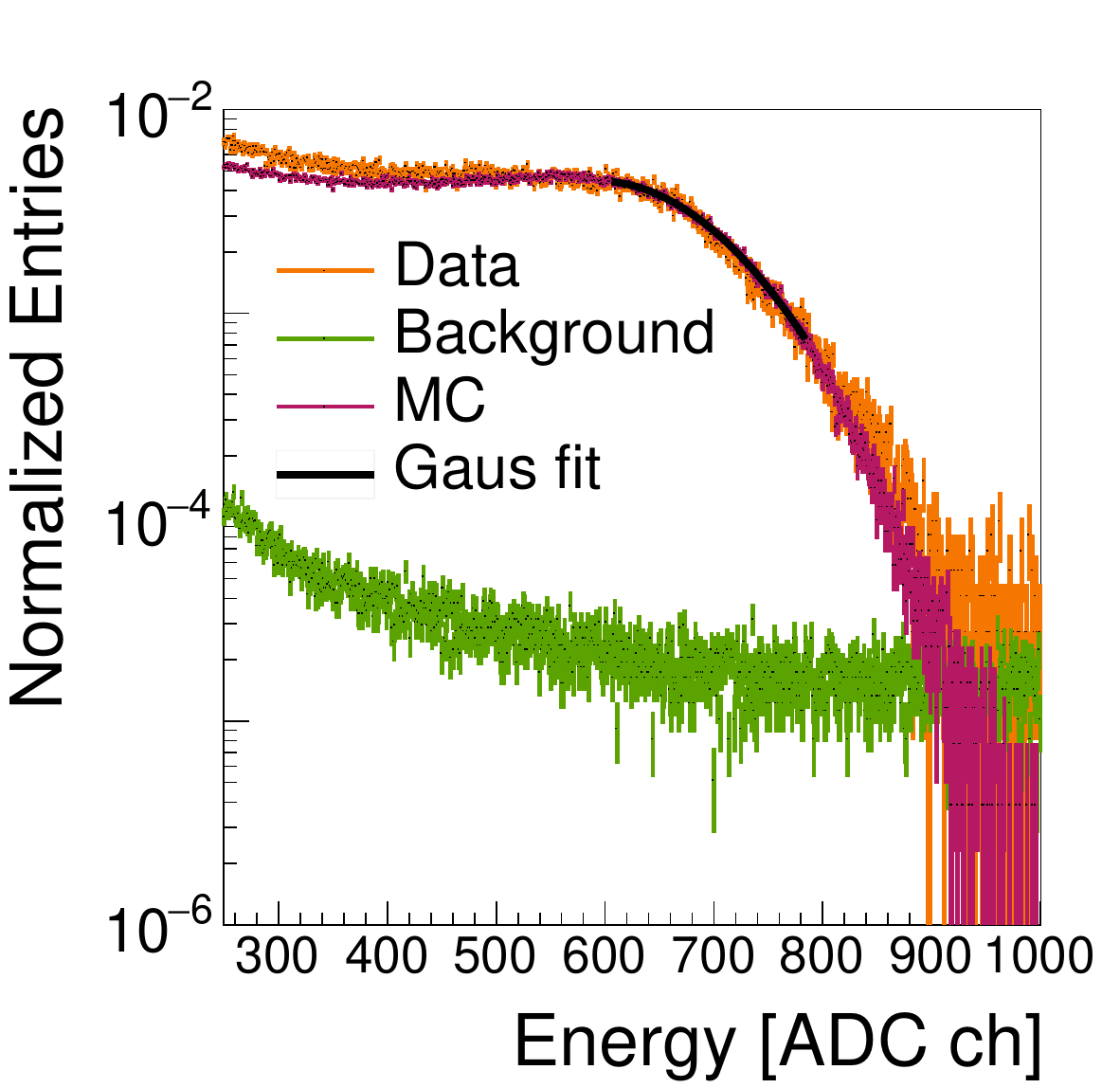}\\
    \end{subfigure}
    \caption{Left: GEANT4 simulation of a \textsuperscript{54}Mn source, folded with different resolution values. 
    Right: Overlay of simulated (red) and measured (orange) \textsuperscript{54}Mn source spectra with a
    15\% resolution applied. The detector background is shown in green and found to be small. The upper half of the edge is fitted with a Gaussian (black line), resulting in an acceptable description with a $\rm \chi^2/ndf$ of 172/178.}
    \label{fig:comptonEdgeRelation}
\end{figure}

To extract the Compton edges from the somewhat featureless calibration source energy spectra, source data are compared to detector simulations.  
The left panel of figure~\ref{fig:comptonEdgeRelation} shows the result of a GEANT4 simulation of the detector response to a $^{54}$Mn point source, folded with various energy resolutions. 

The right panel of figure~\ref{fig:comptonEdgeRelation} shows an overlay of the simulated event distribution, folded with a 15\% resolution, and data. Reasonable agreement between simulation and data is achieved for this resolution. 

As illustrated in the right panel of figure~\ref{fig:comptonEdgeRelation}, a half-Gaussian with centroid $\mu$ and width $\sigma$ is fitted to the spectrum beyond the maximum. 
To determine the edge location from the fit, the data are described by a simple empirical equation, relating the fit parameters with the true location of the Compton edge $\rm C$, inferred from Monte Carlo data: 
\begin{equation} 
\mathrm{C = \mu + \epsilon \cdot \sigma }. \label{eq:ComptonEdge}
\end{equation}
The relational parameter $\epsilon$ is deduced from a set of Monte Carlo simulations of the various energy resolutions and different sources.
This parametrization offers a convenient relation between the spectral shape and the location of the Compton edge, as derived from simulation. 

The energy scale of the scintillation detector is derived using \textsuperscript{54}Mn, \textsuperscript{60}Co, \textsuperscript{22}Na, \textsuperscript{88}Y and \textsuperscript{228}Th source data and simulation. From the simulation of all sources, we derive an average value of $\epsilon=0.77\pm0.08$. A systematic simulation study verified that the value of $\epsilon$ depends only weakly on the chosen energy resolution and $\gamma$-ray energy.
  
The analysis of the energy distributions of all calibration sources shows that the mapping of the ADC channels to energy requires a quadratic relation. 

\subsection{Determination of the acceptance of the event selection cuts}
The \textsuperscript{222}Rn decay rate is inferred from the rate associated with correlated Bi-Po coincidence events. Cuts are applied to the event observables to suppress background and achieve high acceptance. 
Given that the linear dimensions of the liquid scintillation detector are large compared to the range of the charged particles emitted in sequential Bi-Po decays, achievable detection efficiencies can be high. The following section describes how the acceptance of these cuts was determined using radon-source spiked LS.

\begin{figure}
    \centering
    \begin{subfigure}{\textwidth}
        \includegraphics[width=\textwidth]{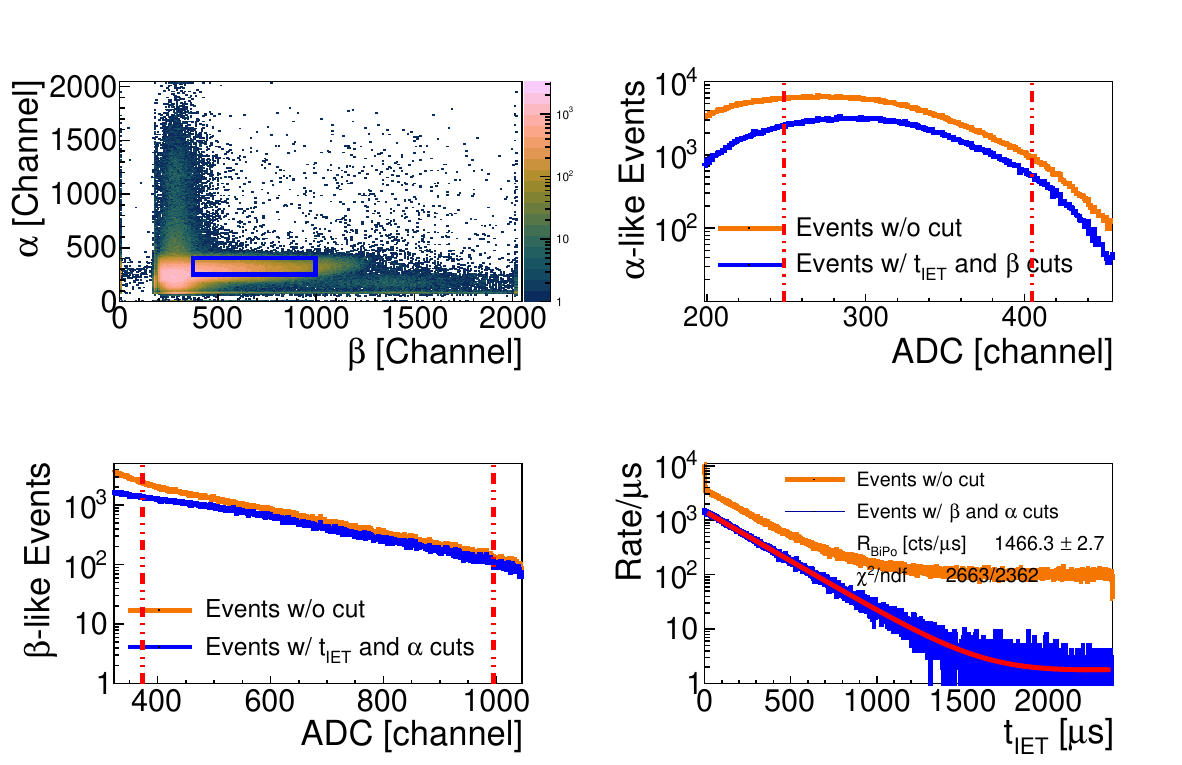}
    \end{subfigure}
    \\
    \begin{subfigure}{\textwidth}
        \includegraphics[width=\textwidth]{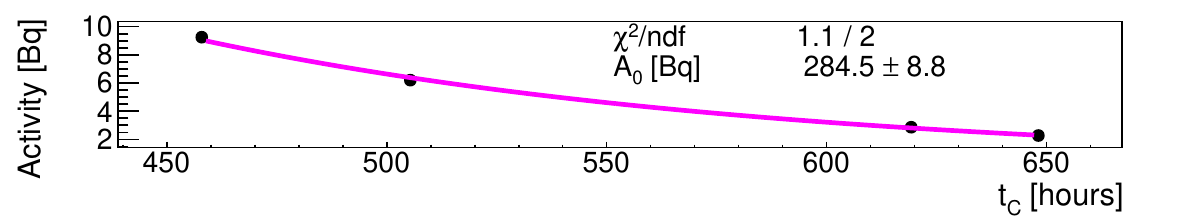}
    \end{subfigure}
    \caption{
    Radon source data collected with the coincidence setup.
    Top left: 2D histogram of the prompt and delayed event energies. The blue rectangular box represents the same energy cuts (red dot-dash lines) shown in the top right and middle left panels. 
    Middle left: 1D histogram of the prompt ($\beta$-like) energy depositions.  
    Top right: 1D histogram of the delayed ($\alpha$-like) energy depositions.
    Middle right: Inter-event time distribution, with the fit superimposed in red. The decay time constant has been fixed to $\rm \tau_{Po}$. 
    Bottom: Repeated Bi-Po measurements of the radon activity of the LS. The time fit, with the decay time constant fixed to $\rm \tau_{Rn}$ is superimposed. Error bars are displayed but are smaller than the markers.}
    \label{fig:optimalCut}
\end{figure}

The \textsuperscript{222}Rn activity of the scintillator contained in the counting cell was determined by Ge detector based $\gamma$-spectroscopy. The cut-acceptances $\rm \varepsilon_{Cut}$ are given by the delayed coincidence rates after cuts divided by the measured activity.

Figure~\ref{fig:optimalCut} shows the energy distributions for the prompt ($\beta$-like, middle left) and delayed ($\alpha$-like, upper right) sub-events for radon spiked scintillator. The vertical red lines indicate the cuts chosen for the interpretation of the data. The orange histogram shows the data without any cuts, the blue with the appropriate analysis cuts applied to the other two observables.
The upper left panel shows an event-density plot in two-dimensional energy distribution. This plot proved useful in identifying runs exhibiting excessive low-energy noise.

The middle right panel of figure~\ref{fig:optimalCut} depicts the histogram of the inter-event time (orange: without, blue: with energy cuts). Except for a 4.5 $\rm\mu$s lower limit cut (corresponding to 1.9\% of \textsuperscript{214}Po decays), no cuts are placed on the time variable. 
The inter-event time histogram is interpreted as the event rate per bin in the inter-event time $\rm t_{IET}$. 
The red line shows the fit of the inter-event time data to an exponential, with its time constant fixed to $\rm \tau_{Po}$, plus a constant random background.
Let the rate of the Bi-Po events, at $\rm t_{IET}=0$, obtained by means of the fit, be denoted as $\rm R_{BiPo}(t_C)$. This fit is repeated for each Bi-Po counting run, with starting at time $\rm t_C$ (with zero chosen to coincide with the end of radon loading) and continuing for the time $\rm \Delta t_{C}$. 
The number of Bi-Po events, observed in each Bi-Po data taking run, is given by integrating the fitted exponential function over $\rm t_{IET}$: $\rm N_{BiPo}(t_C) = R_{BiPo}(t_C) \cdot \tau_{Po}$.

In secular equilibrium (ensured by sufficient waiting before the start of the first Bi-Po run), the number of events collected in each Bi-Po run equals the cut-efficiency-corrected number of radon decays.  
The number of radon events in the time interval starting at $\rm t_{C}$ to $\rm t_{C} + \Delta t_C$ determines the time-dependent radon activity $\rm A_{Rn}(t_C)$. 
We fit this time function to derive the radon activity in LS at time zero as:
\begin{equation}
    \rm A_{Rn,0} = \frac{R_{BiPo}(t_C) \cdot \tau_{Po}}{\varepsilon_{Cut} \cdot \tau_{Rn} \cdot e^{-t_C/\tau_{Rn}}\cdot \left( 1 - e^{-\Delta {t_C}/\tau_{Rn}} \right)}.
    \label{eq:radon_actitivy_0}
\end{equation} 
Note that for typical Bi-Po data taking times $\rm \Delta t_C$ (on average 2.83 days), $\rm \varepsilon_{Cut} \cdot A_{Rn}(t_C) = \varepsilon_{Cut} \cdot A_{Rn,0} \cdot e^{-t_C/\tau_{Rn}}\approx \frac{N_{BiPo}(t_C)}{ \Delta t_C }$ is equivalent to equation~\ref{eq:radon_actitivy_0} to within 12\%.
The lowest panel of figure~\ref{fig:optimalCut} shows the $\rm t_C$ dependence of $\rm A_{Rn}$, as observed in repeated coincidence Bi-Po data-taking runs for a Rn source-loaded sample. 

The magenta fit to an exponential plus constant background, with the time constant fixed to $\rm \tau_{Rn}$, determines $\rm A_{Rn,0}$ and its uncertainty. 
The data shown in the upper panels corresponds to the first, highest statistics, time point in the radon plot (lowest panel). The radon transport model has not been applied at this stage. 

\begin{table}[htb]
\caption[]{Cut efficiencies evaluated for varying energy cuts on the prompt ($\beta$-like \textsuperscript{214}Bi) and delayed ($\alpha$-like \textsuperscript{214}Po) sub-events. Data resulting from the most restrictive cut is shown in figure \ref{fig:optimalCut}. Critical activity (CA) and Minimal detectable activity (MDA) are shown for two cases.} 
\begin{center}
   \begin{tabular}{p{1.5cm} p{2.0cm} p{2cm} p{2.8cm} p{2cm} p{2cm}} 
\hline
 DAQ setup & Prompt energy [keV] & Delayed energy [keV] & Efficiency & CA [$\mu$Bq] & MDA [$\mu$Bq]\\
 \hline
 \hline
 Coinc. & 282 - 2908 & 294 - 997 & 0.624 $\pm$ 0.021 &  & \\
 Coinc. & 386 - 2908 & 359 - 997 & 0.518 $\pm$ 0.018 & &\\ 
 Coinc. & 557 - 2174 & 491 - 862 & 0.288 $\pm$ 0.010 & 444 & 805\\
 Digitizer & < 2094 & 170 - 1195  & 0.689 $\pm$ 0.021 & 454 & 839
 
\end{tabular}
  \label{tab:energycuts}
\end{center}
\end{table}

The acceptance of various cuts was systematically studied with the help of radon-spiked scintillator for both setups. 
Table~\ref{tab:energycuts} shows the acceptances obtained for cuts applied with increasing restrictiveness.
Acceptances in excess of 50\% can be achieved. However, low event-rate runs revealed the intermittent presence of low energy noise events in the coincidence setup surviving the looser energy cuts. 
In order to be able to analyze all data sets collected with this setup with a uniform set of cuts, we decided to adopt the most restrictive energy cuts for all measurements. This approach does not require decisions on the presence of noise events and reduces the reliance on the energy calibration. 
Data collected with the digitizer-based DAQ is free of this intermittent low energy noise. We, therefore, use the loosest cuts for all runs with this setup to benefit from its higher cut efficiency when selecting Bi-Po-coincidence events.

\section{Blank and background determination}
\label{sec:background}

The assay approach described here has three distinct phases. Each of the steps may result in the introduction of unwanted events (background) or radon atoms unrelated to the sample of interest (blank). We classify background and blank into three principal components:
\begin{enumerate}
    \item random coincidences, \label{bkgr:rndm}
    \item radon entering the liquid scintillator in steady state during counting, \label{bkgr:ss}
    \item radon introduced during scintillator handling and cell filling.\label{bkgr:tr}
\end{enumerate}
To understand the different background and blank components, 11 measurements were performed using the full procedure described before but without a sample present in the emanation chamber. We call these runs ``blank runs''. 5 runs with low yield samples were useful in the determination of background components~\ref{bkgr:rndm} and~\ref{bkgr:ss}.
The following sub-sections describe how we quantify and account for these various background components. 

\subsection{Random coincidences background}
Both DAQ systems save sequential events within a time interval 10.6 times the Bi-Po correlation time. Small inter-event times are rich in Bi-Po events, while long ones are essentially due to random coincidences. A fit to the inter-event time distribution is used to quantify the exponential signal and constant background together. The background corrected, time-correlated event rate at time zero and its uncertainty, as seen in the inter-event time fit in figures~\ref{fig:optimalCut} and~\ref{fig:rubberResult}, enter the determination of the radon activity. Random background is, thus, determined together with the sample and accounted for in the data analysis. It only impacts the results via the additional statistical error it causes.

To quantify this background, we fit an exponential plus constant background to the inter-event time distribution, with the time constant fixed to $\rm \tau_{Po}$.
The integral over the constant component, evaluated in a time interval from 4.5 $\mu$s to 3 $\rm \tau_{Po}$, serves as the measure for the random coincidence background rate. 

No significant differences are observed for coincidence setup data taken with either of the two emanation chambers or between 9 blank and 5 low-rate sample runs, taken with the coincidence setup. 
We interpret this to indicate that the random background is associated with the counting setup.
Averaging the random coincidence background
of 14 blank and sample measurements done with the coincidence setup, we obtain: $\rm (25.4\pm 7.3)\cdot 10^{-6} \; counts/s$ (equivalent to 2.2 background events per day). The uncertainty denotes the observed standard deviation of the rate values. The standard deviation is a factor 2.5 larger than the average statistical error of the single run rates, indicating the dominance of other than statistical variability.
The reported random background rates have not been corrected for the radon cut acceptance. 

For 3 blank and background runs with the digitizer setup, we found a higher average random coincidence event rate of $\rm (572\pm 31)\cdot 10^{-6} \; counts/s$. To understand the cause for this substantial rate difference, data was taken with both setups simultaneously and with cells not containing source-related radon. Swapping the signal cables between the different DAQ systems showed that the difference persisted and is not an artifact of the DAQs. Counting of the PMTs and mechanical frames of the two setups with a low background Ge detector shows the digitizer setup to contain more radioactivity than the original coincidence setup.  However, we will show in the next section that the random background has only very little impact on the radon measurement sensitivity. 
This difference is largely inconsequential, as the reported average rate value does not
enter the data analysis.  

\subsection{Steady state radon blank}
Radon atoms, decaying in the liquid scintillator but unrelated to the sample under study, can enter the liquid scintillator due to cell components outgassing or small gas leaks. As in the case of the random background, we are using a time analysis to determine this background component during sample counting. This is done by measuring the correlated event rate in repeated Bi-Po runs, at $\rm t_C$ much larger than $\rm \tau_{Rn}$. The time fit to the radon decay curve, shown in the bottom panel of figure~\ref{fig:rubberResult}, separates the sample-related transient (exponential) radon activity from the steady-state (constant) blank radon activity. This approach allows to determine the blank individually for each sample measurement. The added selectivity comes at the expense of a counting time lasting up to 3 weeks per sample. This time, together with the long radon accumulation time, determines the achievable sample throughput of the method.

Averaging over the 9 blank runs, taken with the coincidence setup, we observe that the steady state radon activities of both setups are compatible with each other after correcting for their different cut efficiencies. We, thus, report this background in terms of a cut-corrected $^{222}$Rn activity. We further do not observe significantly different rates when averaging the 9 blank and 5 low-rate sample runs.
Again, we conclude that this background is caused by the setup itself, especially the counting cells. Averaging over all data runs, we observe an average steady state $^{222}$Rn decay rate of $\rm 126\pm 153\; \mu Bq$ in the LS (equivalent to 3.1 background events per day). We quote the standard deviation determined over all runs and note that it is a factor of 3.4 larger than the average run-wise statistical error.
 
This blank is uncritical, with little impact on the sensitivity, as it is determined from the sample data of each measurement. The reported average blank radon activity value does not enter into the data analysis.

\subsection{Handling related radon blank}
Radon contained in the carrier gas, outgassing of the emanation chamber and its piping, radon contained in the liquid scintillator because of small leaks in the storage container, and radon introduced during handling and filling of the scintillation cell needs to be quantified. The decay of this radon ``blank'' forms a transient component, just like the sample-related radon. It cannot be determined from the time distributions and needs separate measurements without sample. Because of the dependence on conditions and handling, we decided to assess the variability from 11 repeated blank measurements with both setups and not assume the dominance of statistical variability. It is the most challenging and time consuming to determine. 

Analyzing the 11 blank runs, we observe no significant difference between the two emanation chambers nor between the two setups. To combine the data of both setups, cut efficiencies have been corrected. The average transient $^{222}$Rn activity and standard deviation of the LS, determined from the 11 blank runs, are $\rm 423\pm 288\; \mu Bq$ (equivalent to 10.5 events/day).  
As this measurement involves gas transport, we also report the results in terms of a radon outgassing blank. After application of the loading model, the blank rate becomes $\rm \mathcal{R}_{blank}=136\pm 92$ Rn atoms/day.

\subsection{Minimal detectable activity}
Random coincidence and steady-state radon background rates are determined together with the sample-related rate from their different time dependencies. 
Thus, no separate subtractions are needed. 

The analysis of sample outgassing data does need to account for the radon introduced by handling, $\rm \mathcal{R}_{blank}$.
To get the net outgassing result of a sample run,
$\rm \mathcal{R}_{blank}$ needs to be subtracted from the observed rate, $\rm \mathcal{R}_{S}=\mathcal{R}-\mathcal{R}_{blank}$.
To estimate the uncertainty introduced by this subtraction, we evaluated two approaches under the null hypothesis. 
In the first approach, $\rm \mathcal{R}_{blank}$ is fixed to the average value obtained in the 11 blank measurements. 
The uncertainty is then estimated by propagating the standard deviation of $\rm \mathcal{R}_{blank}$ with the individual standard deviation of each randomly drawn dataset. 
In the second approach, we subtracted a blank rate drawn from a Gaussian distribution, characterized by $\rm \mathcal{R}_{blank}$ and its standard deviation. 
While the second method best reflects reality, it cannot be performed for each measurement. 
We compare the two methods to justify the approximate approach.  

Using the first method, we find a mean and RMS of 48 and 310 $\rm \mu$Bq, respectively. In the second approach, these values were 5 and 390 $\rm \mu$Bq, respectively. 
Given the large standard deviation, both methods are found to be essentially bias-free. 
We attribute the difference between the two to deviations from Gaussianity. We find the result of the approximate approach to be acceptable.

\begin{figure}
    \centering
    \begin{subfigure}{0.49\textwidth}
        \centering
        \includegraphics[width=\textwidth]{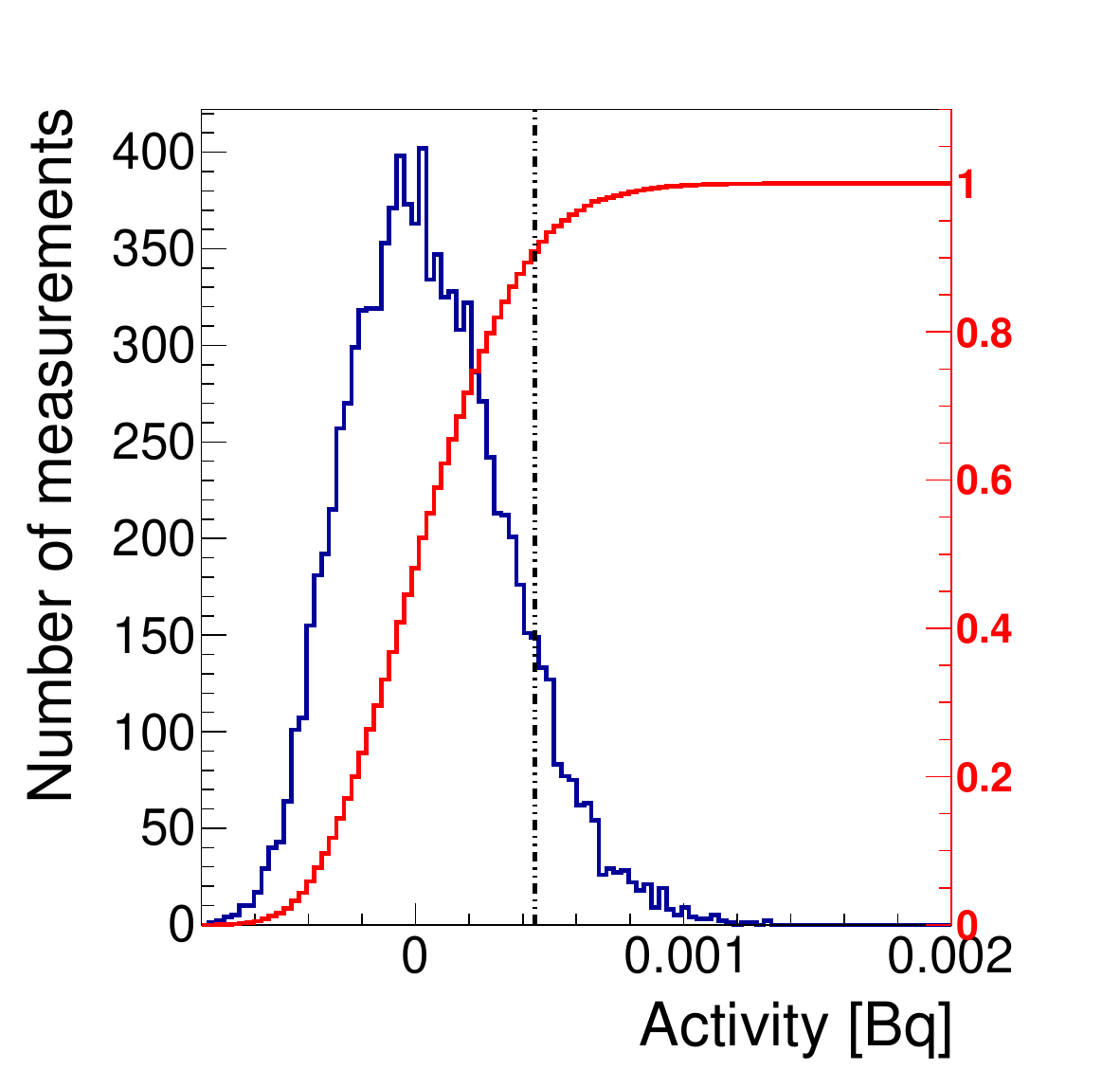}\\
    \end{subfigure}
    \begin{subfigure}{0.49\textwidth}
        \centering
        \includegraphics[width=\textwidth]{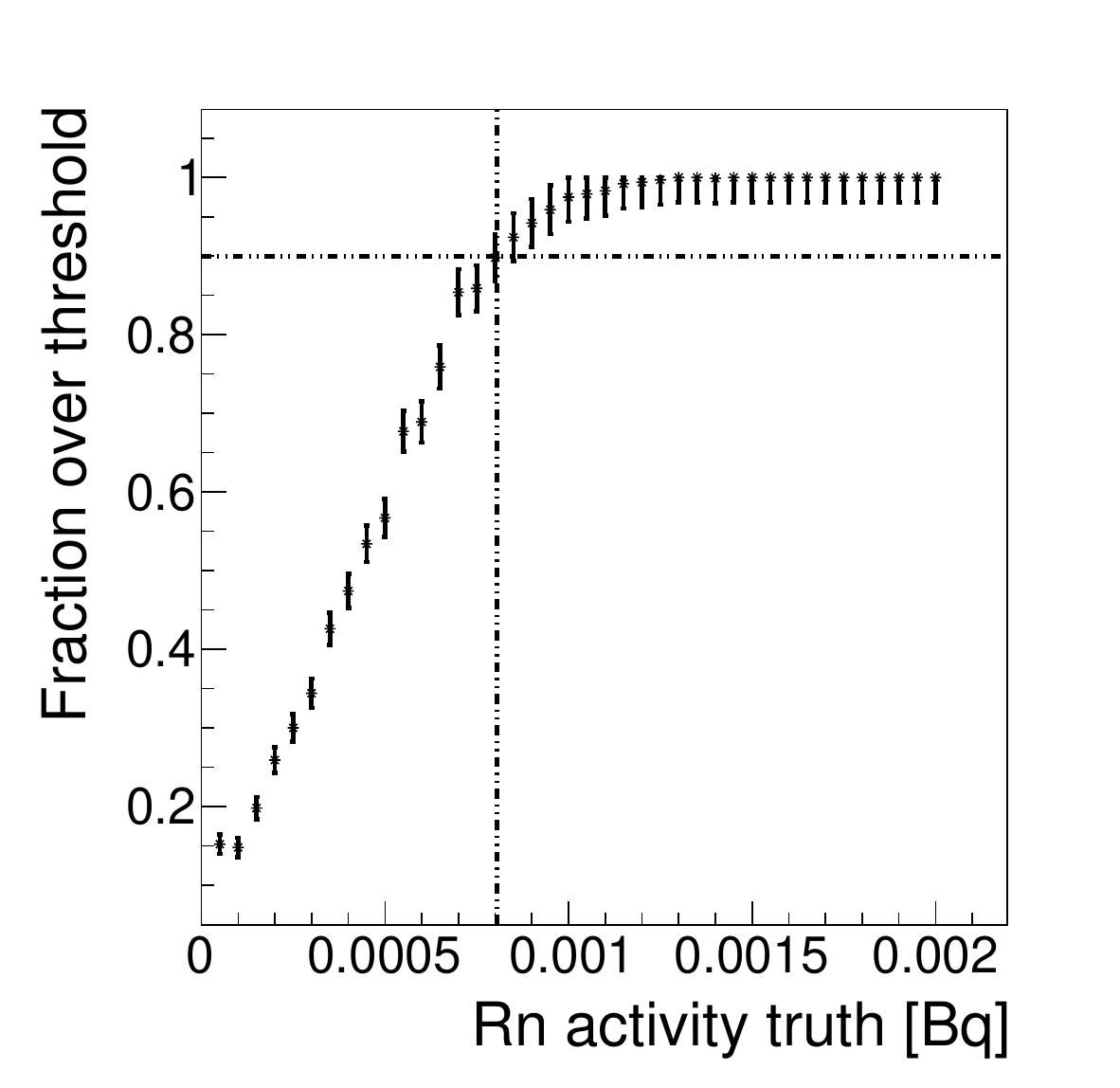}\\
    \end{subfigure}
    \caption{Left: frequency distribution (blue) of 10,000 radon activities, returned by the fit under the null hypothesis. The data-derived blank event rate has been subtracted. A critical activity of $\rm 444\; \mu Bq$ is determined from the 90\% percentile of the plotted distribution, (indicated by the vertical dash-dotted line). The red curve is the corresponding cumulative distribution. 
    Right: a minimal detectable activity of $\rm 805\; \mu Bq$ is determined as the lowest radon activity with 90\% of their blank-subtracted values above the critical activity. The error bars correspond to the Poissonian error of the number of events contained in the tail above the critical activity. 
    }
    \label{fig:sensitivity}
\end{figure}

We define a critical activity (CA) of the radon setup to describe our ability to distinguish a sample-related counting rate from blank and background.
$\rm \mathcal{R}_{CA}$ is determined, under the null hypothesis, as the smallest activity with a fraction of false positives not exceeding the confidence level, as shown in figure~\ref{fig:sensitivity}.
 
Net rates exceeding the critical activity 
of the setup: $\rm \mathcal{R}_{S}>\mathcal{R}_{CA}$ are reported as observations, made at a certain confidence level. Measurements not satisfying this condition are reported as $\rm <\mathcal{R}_{MDA}$ (defined below), corresponding to the minimal detectable activity (MDA), again at a certain confidence level. 
$\rm \mathcal{R}_{CA}$ and $\rm \mathcal{R}_{MDA}$ were determined by means of toy Monte Carlo simulations. 

10,000 repeated radon analyses were simulated, using the blank measurement-defined means and standard deviations of all three types of background. On average, each measurement, sample or blank, consists of 5.1 repeated Bi-Po counting runs, each lasting, on average, 2.83 days.
The CA estimate assumes that a radon measurement consists of 5 Bi-Po counting runs with an equal duration of 2.83 days.
 
To achieve appropriate randomization, each measurement used a single random realization of each background rate probability density function (PDF), assumed to have Gaussian characteristics. The resulting random draws of the background rate are then in common to all five Bi-Po counting runs. Each new radon run received new random rate values.
The number of events in each run was based on a Poissonian randomization of that particular rate. 
Random coincidence events were distributed uniformly over the inter-event time distribution. Steady state and transient radon background received the appropriate exponential Bi-Po $\rm t_{IET}$ distribution. The transient radon blank rate was diminished according to $\rm \tau_{Rn}$ for the 5 successive Bi-Po analyses while the other two backgrounds were assumed to be constant in $\rm t_C$. Before fitting, all backgrounds were superimposed on each other.

The analysis of the Monte Carlo data was done with the same double time fit as used for detector data.
The blank subtracted result of 10,000 repetitions of this procedure is shown in the left panel of figure~\ref{fig:sensitivity} for the null hypothesis. 68, 90 and 95\% quantiles of 175, 444 and 571 $\mu$Bq were found for the $^{222}$Rn activity in LS, respectively. Assuming the use of small samples (all of the emanation chamber is filled with gas), a 48 min bubbling time, a 20 ml/min gas flow rate and $\rm V_{LS}=175\; ml$, these CAs can be converted into radon outgassing rates of $\rm \mathcal{R}_{CA}=56,\; 142\; and\; 183\; ^{222}Rn\; atoms/day$ at the above confidence levels, respectively.

To determine which background dominates the CA threshold, simulations were performed by setting two of the three to zero and not subtracting the blank so as not to introduce biases. The CAs derived for the random coincidence and the steady state radon background are about a factor of 10 and 5 smaller than the combined threshold. This shows that the CA is largely limited by the radon introduced during accumulation, transfer and handling and not the counting setup.

The determination of $\rm \mathcal{R}_{MDA}$ also needs to account for false negatives and, therefore, the statistical variability of a hypothetical signal. To implement this additional variability, varying radon activities were added to the simulated background. Random draws over its Poissonian fluctuation (in terms of integer counts) were used for every Bi-Po run. 
The right panel of figure~\ref{fig:sensitivity} shows the fraction of radon activity values returned above the CA. Using 68, 90, and 95-percentiles results in MDAs of 564, 805 and 924 $\mu$Bq, respectively. The 90-percentile is indicated by the dash-dotted lines.

These MDA limits correspond to outgassing rates of $\rm \mathcal{R}_{MDA}= 181,\; 258,\; 296 \; atoms/day$. 
The simulations done with known net sample activities further show the analysis to be largely bias free.

\section{Validation}
\label{sec:validation}
\begin{figure}
    \centering
    \begin{subfigure}{\textwidth}
        \includegraphics[width=\textwidth]{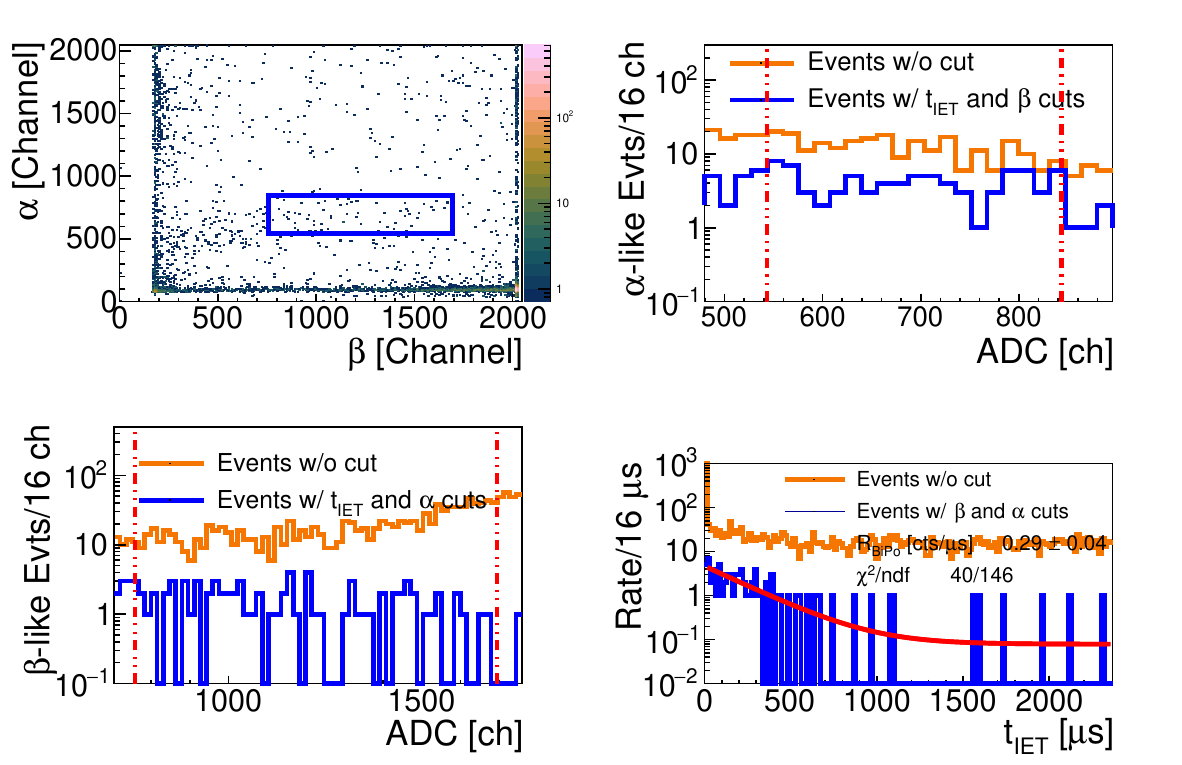}
    \end{subfigure}
    \\
    \begin{subfigure}{\textwidth}
        \includegraphics[width=\textwidth]{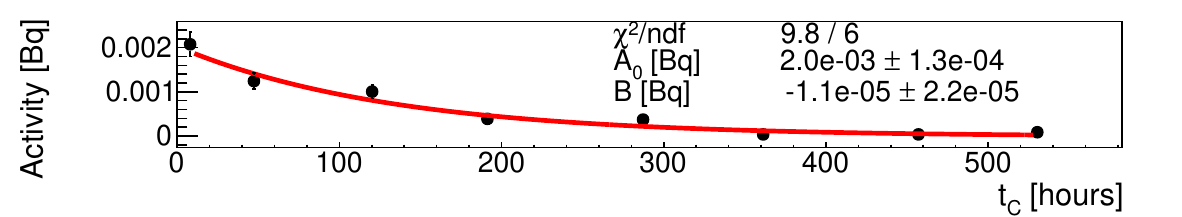}
    \end{subfigure}
    \caption{Radon-emanation data from butyl rubber sample. The same cuts as in figure~\ref{fig:optimalCut} were applied. To reduce fluctuations, the inter-event time histogram has been rebinned by a factor 16 compared to the data in figure~\ref{fig:optimalCut}. Note the fitted radon activity, $\rm A_0$, shown in the bottom panel, is not corrected for the handling blank and liquid scintillator transfer fraction from the bubbler to the cell.}
    \label{fig:rubberResult}
\end{figure}

In order to validate the techniques described in this paper, the outgassing of a standard of sorts was evaluated. 
The determination of CA and MDA made use only of statistical uncertainties. When reporting radon outgassing rates of samples, the systematic uncertainty of the model is stated separately. This needs to be done for both activity and outgassing rate to properly account for differences in $\rm V_C,\; V_{LS},\; \phi,; t_L$.

To have a sample resulting in an emanation rate well above $\rm \mathcal{R}_{CA}$ but still ``small'' compared to source data, we chose a $\rm 61\; cm\,\times\, 7.5\; cm \,\times\, 0.32\; cm$ piece of butyl rubber supplied by SNOLAB / Laurentian University. Multi-lab radon counting results for this sample were reported in reference~\cite{LZradioactivityPaper}, making it a candidate for a consistency check. The SNOLAB / Laurentian University counting result used in~\cite{LZradioactivityPaper} differs from our new measurement of $\rm 0.542\pm 0.11^{stat} \pm 0.078^{syst}\; atoms/(day \cdot cm^2)$, by 0.8 standard deviations. The measured rate corresponds to $\rm A_{LS} = 1760 \pm 340^{stat} \mu Bq$ (assuming equal loading efficiency).  We consider the observed outgassing rate that is 20\% larger than the reference value close enough to serve as verification of functionality at small activities. The corresponding data is presented in figure~\ref{fig:rubberResult}.

\section{Conclusion}

Radon emanation measurement based on Bi-Po delayed coincidence counting has its strengths and weaknesses. 
This study shows how coincidence counting and long observation times can differentiate radon decay from some of the background. We show that radon introduced during handling limits the critical activity. Critical and minimal detectable activities are determined by Monte Carlo. We show that detection efficiencies as high as 70\% can be achieved. However, the radon transfer and loading efficiency of about 25\%, based on a model described in the paper, ultimately limits the minimal detectable activity of the method.

\acknowledgments
We thank our colleagues from the LZ and nEXO collaborations for their support and many fruitful discussions.
This research was supported, in part, by the U.S. Department of Energy under DOE Grants DE-SC0012447 and DE-FG02-01ER41166. 
We acknowledge the early contributions of Tamar Didberidze to this work and thank our colleague Jacques Farine for providing us with the reference rubber sample and a measurement of its radon outgassing.

\bibliographystyle{JHEP}
\bibliography{biblio.bib}
\end{document}